\documentclass[superscriptaddress,aps,preprint,longbibliography]{revtex4-2}
\usepackage{graphicx} 
\usepackage{amsmath}
\usepackage{amssymb}
\usepackage[utf8]{inputenc}
\usepackage{booktabs}
\usepackage{enumitem}
\usepackage{xcolor}
\graphicspath{{./pictures/}}
\begin{document}
\title{On early-warning of full versus partial Atlantic overturning circulation collapse}
\author{Johannes Lohmann}
\affiliation{Physics of Ice, Climate and Earth, Niels Bohr Institute, University of Copenhagen, Denmark}

\begin{abstract}
Climate models indicate a significant slowdown of the Atlantic Meridional Overturning Circulation (AMOC) in the future, and some suggest it may collapse irreversibly to a substantially weakened state. The global warming threshold where this may happen is highly uncertain. An alternative to model-based threshold estimation are early-warning signals (EWS) in AMOC fingerprints, which can predict destabilization of a steady state (a saddle-node bifurcation) from generic changes in statistical properties. But an AMOC collapse may be a sequence of partial weakenings with shutdown of deep water formation in distinct regions.

A conceptual model featuring sequential tipping points in two such regions is presented. Since the system only follows the expected saddle-node normal form when very close to each tipping point, a variety of trends in EWS are seen for different observables. This makes it hard to determine what type of collapse (partial or full) will follow, and when it will happen.
\end{abstract}

\maketitle

\section{Introduction}

The Atlantic Meridional Overturning Circulation (AMOC) is a major climate pattern  and a large contributor to the global meridional heat transport \cite{JOH23}. Increased warming (polar amplification) and freshening of northern Atlantic surface waters could lead to a shutoff of convection and thereby the AMOC's current downwelling path. Due to the likely existence of the salt-advection feedback \cite{RAH96,ARU24} - whereby an initial AMOC weakening would be amplified in a self-reinforcing way by decreasing surface salinity in downwelling regions - this may happen abruptly and irreversibly. An AMOC collapse would lead to dramatic decreases in Northern Europe temperatures while warming the Southern Hemisphere and producing various cascading impacts \cite{JAC15}.

A significant AMOC weakening is still hard to determine, because (besides sparse snapshot measurements \cite{BRY05,KAN10}) direct monitoring only covers the past two decades \cite{JOH23,LOZ23}. An indirect extension of the record back in time via sea level anomalies indicates a downward trend \cite{FRA15}, but other methods based on boundary density anomalies and air-sea heat fluxes do not find a significant trend \cite{WOR21,TER25}. Much longer proxy reconstructions based on sea surface temperature (via its sensitivity to AMOC heat transport) suggest an AMOC weakening in the industrial era \cite{RAH15,THO18,CAE18,CAE21}, but the fidelity of such fingerprints remains debated \cite{MOF19,KIL22,CAE22,LAT22}.

\begin{figure}
\includegraphics[width=0.99\textwidth]{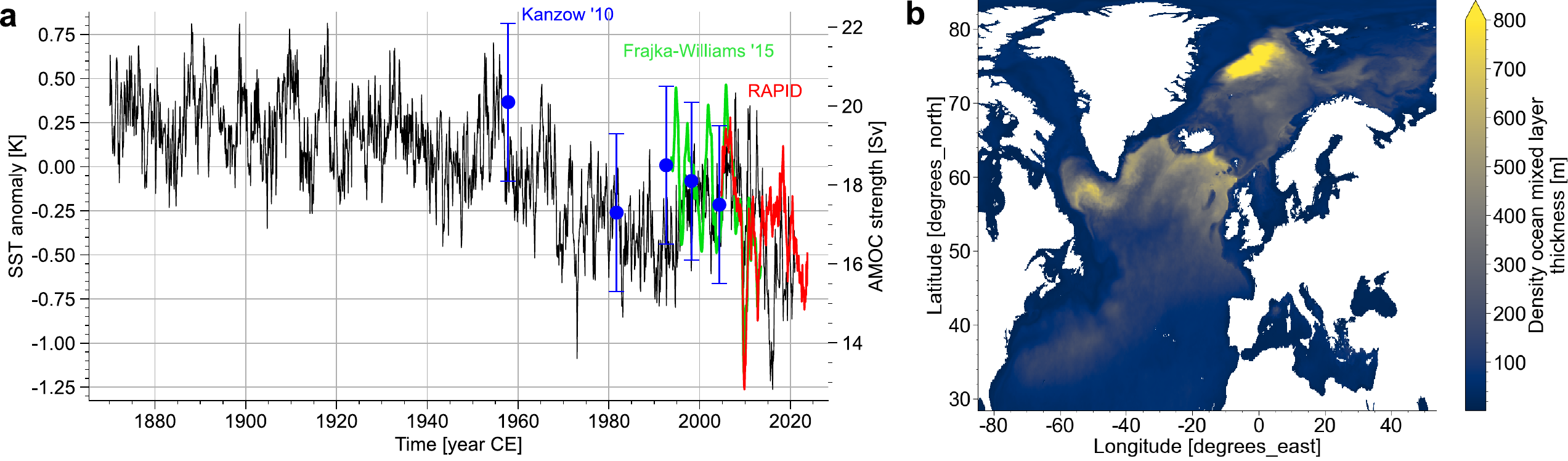}
\caption{\label{fig:fig1}
{\bf a} Time series of observed and reconstructed changes in the AMOC circulation strength.
Included are direct AMOC strength measurements from the RAPID array (red) and five hydrographic sections (blue dots) \cite{BRY05,KAN10}, as well as a reconstruction based on sea level anomalies (green) \cite{FRA15}.
The longest time series (black) is the sea surface temperature fingerprint \cite{RAH15}
sometimes used as proxy for AMOC strength. Shown is the anomaly of sea surface temperatures  in the North Atlantic subpolar gyre with twice the global mean sea surface temperature anomaly to correct for global warming and polar amplification, as was used in \cite{DIT23}.
{\bf b} Climatology of the mixed layer depth for the month of March from the GLORYS reanalysis (period 1993 to 2024).
}
\end{figure}

Most climate model projections until the end of this century show a significant AMOC decline \cite{WEI20} although differing in its magnitude \cite{BON25}. Models with extended simulations until the year 2300 or longer can transition to a state where North Atlantic deep water formation ceases and the AMOC becomes shallow and weak, with downwelling restricted to the subtropics \cite{ROM23,DRI25}. For some models this can happen even under low emission scenarios \cite{DRI25}.
Increased freshwater forcing from Greenland melt \cite{BAK16,HOF20} can also weaken the AMOC, but is missing in most projections. Models have shown AMOC collapses at meltwater rates of 0.2~Sv \cite{HAW11}, 0.3~Sv \cite{JAC23}, 0.5~Sv \cite{VWE23,VWE25}, or as low as 0.125~Sv in a recent eddy-resolving ocean-only simulation \cite{VWE25b}.
An average and peak discharge of 0.07~Sv and 0.16~Sv, respectively, was estimated in a complete Greenland melt under high-emission temperatures \cite{ASC19}. Since at 0.3~Sv (0.1~Sv) all of the 2.74 mio Gt Greenland ice would be melted after 290 (870) years, it is uncertain whether high enough melt will be sustained long enough to complete an AMOC shutdown. But the combined freshwater and global warming (or atmospheric CO$_2$) thresholds of the AMOC is not known.

Overall, the possibility of an AMOC collapse under plausible future emission scenarios cannot be discarded \cite{DIJ26}. It is difficult to give model-based quantitative estimates of AMOC tipping thresholds, and the observational AMOC record is too short to indicate a significant trend, which from a dynamical systems view may also evolve non-linearly or increase before a collapse \cite{SOO25}. It is thus relevant to consider an independent line of evidence by monitoring changes in AMOC variability that may show generic precursor signals of destabilization.
These can be measured by statistical early-warning signals (EWS) due to critical slowing down (CSD) before bifurcations \cite{WIS84,SCH09}. If the AMOC system is a dynamical system forced by stochastic atmospheric perturbations and driven towards a saddle-node bifurcation by anthropogenic emissions, an impending AMOC collapse should be detectable by increasing variance and autocorrelation of fluctuations of an observable that measures the relevant AMOC fluctuations.

This has been attempted before with the conclusion that the AMOC is indeed approaching a TP \cite{BOE21a,MIC22,DIT23}, but since direct measurements are too short, fingerprints (e.g. Fig.~\ref{fig:fig1}a, black time series) must be used. Besides assumptions on memory \cite{MOR24b}, stationarity \cite{BEN23,MOR24} and state-dependence \cite{MOR23} of the noise process that need to be verified, it is debated how well the fingerprints capture AMOC variability \cite{TER25}. A whole spectrum of fingerprints could be potentially used, and thus inferred trends in EWS come with an uncertainty that is hard to quantify. An analysis of historical variability in sea surface temperature or salinity shows EWS in North Atlantic regions with a plausible AMOC fingerprint, but also nearby locations where variability decreases \cite{BEN23}. It is thus not straightforward to attribute increased variability in selected regions as clear sign of CSD.

Since the correct fingerprint to use for EWS is unknown, simulated AMOC collapses have been analyzed for variables and locations with increasing variance and autocorrelation. An analysis of the FAMOUS model yielded variability increases of the overturning streamfunction at specific latitudes \cite{BOU14} - although not the same ones for variance and autocorrelation - or increased coherence of fluctuations in the deep ocean \cite{FEN14}. Other models show no EWS in the AMOC strength before a collapse \cite{VWE24,LOH25}. An analysis of a quasi-equilibrium AMOC collapse in the CESM model showed that salinity in the deep Atlantic at its southern border would be an ideal place to observe increased variability \cite{SMO25}. However, variance and autocorrelation show very different behaviour and there are equally many regions with decreases in variability. Another analysis of the model showed increased variability in Nordic sea surface temperatures, however other North Atlantic regions often associated with an AMOC fingerprint feature decreased variability \cite{PAT26}. It is thus unclear which of these signals is a robust CSD signal, and whether other models would show the same. Variance and/or autocorrelation may increase for reasons other than CSD, such as changes in deterministic, oscillatory modes \cite{LOH24}.

It is not guaranteed that changes in variability are due to CSD when a) finding an observational fingerprint for the AMOC strength and use it for EWS, or b) when searching variables and locations with highest variability increase leading up to a simulated AMOC collapse. Instead, specific observables need to be chosen for EWS via physical reasoning on the suspected feedback processes \cite{VWE24,PAT26} or on dynamical systems grounds \cite{LUC24,LOH24b,LOH25b}. The latter may be obtained by knowledge of a model edge state \cite{LOH24b,LOH25} or by data-driven operator methods \cite{LOH25b}.

The purpose of this paper is to illustrate additional difficulties in interpreting EWS when it is unknown whether the AMOC collapses in a sequence of one or more tipping points, and if various observables are available.

The usual interpretation of EWS for an AMOC collapse is in analogy to the Stommel model \cite{STO61}. It shows a bistable AMOC as a result of the positive salt-advection feedback, whereby as atmospheric temperature or freshwater forcing changes, a saddle-node bifurcation occurs where the present-day state loses stability and the circulation enters a weak salinity-driven state.
A wide range of other conceptual models have been proposed to improve on the physical limitations of the Stommel model, for instance that it does not explicitly represent vertical mixing in the North Atlantic \cite{WEL82}, and that it only covers one Hemisphere \cite{ROO82}. This was often done by adding further zonally averaged boxes, which can give rise to additional solutions (e.g. mid-latitude sinking) \cite{HUA92} or bifurcations into oscillating regimes that may be relevant for abrupt past climate changes \cite{WIN93, COL06, STA07, NAD25}.
Other extended box models show that the AMOC collapse may occur due to a homoclinic bifurcation followed by a subcritical Hopf bifurcation \cite{SCO99,WOO19,ALK19},
which is also relevant for the interpretation of EWS (not further discussed here).

In this paper, a different extension to the Stommel model is considered, which illustrates the possibility of spatially distributed downwelling paths. In the model, a well-mixed subtropical Atlantic ocean box is connected to two distinct but coupled northern Atlantic boxes with heterogeneous forcing, representing the Irminger-Labrador and Nordic seas, respectively.
This is motivated by the idea that an AMOC collapse may be initiated locally, e.g., in the Labrador sea, and by a positive feedback other than the salt-advection feedback, such as Welander's convective \cite{WEL82} or subpolar-gyre feedbacks \cite{BOR13,DIJ26}.
Figure~\ref{fig:fig1}b shows how deep mixed layers occur across multiple locations in the Labrador and Irminger seas, the Icelandic basin, as well as the Nordic seas. This suggests that deep water formation is distributed across regions, which contribute with different amounts to the AMOC downwelling and return path \cite{PET20,LOZ23,ART23}. The regions are subject to different evolving conditions: There is larger Greenland meltwater input in the Irminger-Labrador seas compared to the Nordic seas \cite{BAM12}, and larger heat loss in the Nordic seas \cite{CHA19}.
Hence, a local shutdown of convection may happen at different global warming levels.

It can thus be questioned whether an AMOC collapse will be a result of a single TP corresponding to a saddle-node bifurcation. Indeed, prior to an AMOC collapse there can be further intermediate TPs related to the subpolar gyre \cite{SGU17,SWI21,LOH24} or changes in spatio-temporal patterns of the flow \cite{LOH24}. As a result of such potential sequences of tipping events, EWS indicators may exhibit several jumps with non-monotonic behaviour in between \cite{LOH24}. By extension, a monotonic EWS trend indicative of a TP generally does not inform whether it will be a full AMOC collapse, a localized transition, or a change in spatio-temporal mode.
It also follows that EWS-based extrapolations to predict the time of tipping \cite{DIT23} are highly dependent on the observable used, since the required scaling of variance or autocorrelation only arises when the observable obeys the saddle-node {\it normal form}, which may not happen until arbitrarily close to the TP.

The conceptual model presented here is intended to highlight these difficulties in interpreting EWS prior to a suspected AMOC collapse. The emphasis is on the issue of distinguishing from EWS whether there will be a complete or partial collapse,
given that there is no a priori consensus on whether the AMOC would collapse fully in one bifurcation, via a sequence of bifurcations, or gradually with no bifurcations at all.
The model inherits the main limitations of Stommel's model, and does not attempt to give a more realistic mechanism for the AMOC or the likelihood of its collapse. Instead, it serves as a minimal extension that yields a more complex multistability owing to the spatial heterogeneity.

The structure of the paper is as follows. In Sec.~\ref{sec:model} the model and its bifurcation structure is presented. Sec.~\ref{sec:ews} discusses the EWS found across observables for different TPs in the model. The possibility for quantitative prediction and its dependence on observables and a partial versus a full collapse is analyzed in Sec.~\ref{sec:prediction}. Discussion and conclusions are given in Sec.~\ref{sec:discussion}.

\section{AMOC box model with two convection sites}
\label{sec:model}

\begin{figure}
\includegraphics[width=0.5\textwidth]{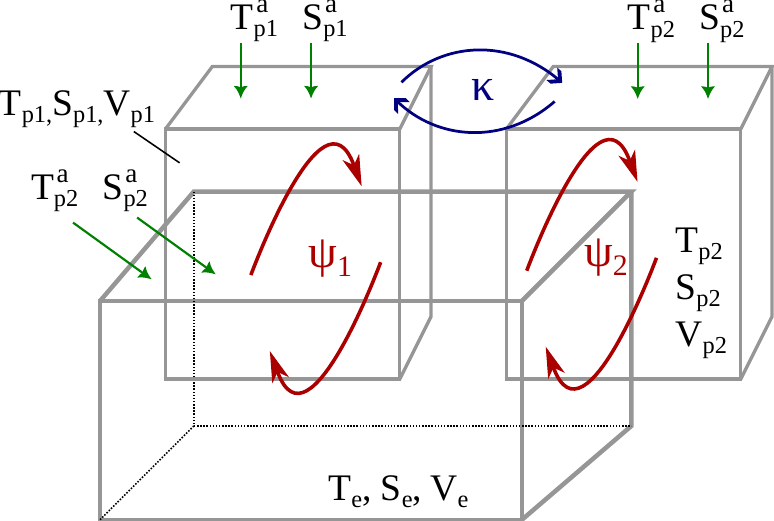}
\caption{\label{fig:box_model} 
Schematic of the three-box AMOC model. In the foreground is the larger equatorial box, and in the background are two northern polar boxes. Bidirectional coupling by the overturning flow $\Psi_1$ and $\Psi_2$ between the boxes is represented by the red arrows. The direction of the arrows illustrates the present-day, temperature-driven AMOC 'ON' state, where the density of the polar boxes is larger than in the equatorial box ($q_1>0$ and $q_2>0$). As in Stommel's model \cite{STO61}, a ``reversed'' (in terms of the meridional density gradient) and weaker salinity-driven circulation state can exist for either of the polar boxes, in which case $q_i<0$ and the arrows to and from a given polar box are reversed. If the circulation is reversed for both polar boxes it corresponds to a fully collapsed AMOC 'OFF' state. If only reversed for one polar box it is a partially-collapsed (PC) state.
}
\end{figure}

A schematic of the conceptual AMOC model is shown in Fig.~\ref{fig:box_model}. It represents a well-mixed equatorial box connected by the overturning flow to well-mixed polar basins. Here, two polar boxes with different atmospheric forcing represent the Labrador-Irminger and Nordic seas.
In each box $X$ , temperature and salinity relax to prescribed atmospheric forcing values $T_X^a$ and $S_X^a$, with relaxation time scale proportional to the box volume $V_X$. Temperature and salinity in the polar boxes are coupled linearly with strength $\tilde{\kappa}$. This coupling is a parsimonious representation of zonal exchange mainly via the wind-driven circulation. The polar boxes are coupled to the equatorial one by the overturning flow $\Psi(\Delta \rho_{1,2})$, which depends on the meridional density gradient between the equatorial and each of the polar boxes $\Delta \rho_{1,2} = \rho_{p1,2} - \rho_e$. The resulting governing equations of temperature and salinity in the polar box 1 are
\begin{align}
\begin{split}
\label{eq:polar}
V_{p1} \dot{T}_{p1} &= V_{p1} R_T (T_{p1}^a - T_{p1}) + \Psi(\Delta \rho_1)(T_e - T_{p1}) + \tilde{\kappa} (T_{p2} - T_{p1}) \\
V_{p1} \dot{S}_{p1} &= V_{p1} R_S (S_{p1}^a - S_{p1}) + \Psi(\Delta \rho_1)(S_e - S_{p1}) + \tilde{\kappa} (S_{p2} - S_{p1}).
\end{split}
\end{align}
A quadratic dependency of flow strength and meridional density gradient is chosen, 
as proposed by Cessi and Young \cite{CES92,CES94}. It is different from the absolute value function used by Stommel \cite{STO61}, but gives qualitatively similar dynamics, while avoiding non-smooth dynamics and bifurcations. This yields
\begin{equation}
 \Psi(q_1) = \gamma \rho_0^{-2}(\rho_{p1} - \rho_e)^2 = \gamma (\alpha_T(T_{p1} - T_e) - \alpha_S(S_{p1} - S_e))^2
\end{equation}
where the latter equality follows from a linear equation of state with respect to a reference state (subscript '0')
\begin{equation}
 \rho_X = \rho_0(1 - \alpha_T(T_X - T_0) + \alpha_S(S_X - S_0).
\end{equation}
The equations for polar box 2 are the same as Eq.~\ref{eq:polar} when exchanging indices '$p1$' and '$p2$'. The equatorial box is goverened by 
\begin{align}
\begin{split}
V_e \dot{T}_e &= V_e R_T (T_e^a - T_e) + \Psi(\Delta \rho_1)(T_{p1} - T_e) + f(q_2)(T_{p2} - T_e)  \\
V_e \dot{S}_e &= V_e R_S (S_e^a - S_e) + \Psi(\Delta \rho_1)(S_{p1} - S_e) + f(q_2)(S_{p2} - S_e)
\end{split}
\end{align}
By subtracting the equations of the equatorial box from the polar ones, closed equations are obtained that only depend on the meridional gradients in temperature $T_1 \equiv T_{p1} - T_e$ and salinity $S_1 \equiv S_{p1} - S_e$.
To reduce the number of parameters, equal volumes of the polar boxes $V_{p1} = V_{p2} \equiv V$ are assumed, and $V_e \equiv \tilde{\omega} V$ is defined. After rescaling time by $R_T$, this yields
\begin{align}
\begin{split}
\dot{T}_1 &=  \Delta T_1^a - T_1 - \frac{\Psi(\Delta \rho_1)(1 + \tilde{\omega})}{\tilde{\omega} V R_T} T_1 + \frac{\tilde{\kappa}}{V R_T} (T_2 - T_1) - \frac{\Psi(\Delta \rho_2)}{\tilde{\omega} V R_T} T_2 \\
\dot{S}_1 &= \frac{R_S}{R_T} (\Delta S_1^a - S_1) - \frac{\Psi(\Delta \rho_1)(1 + \tilde{\omega})}{\tilde{\omega} V R_T} S_1 + \frac{\tilde{\kappa}}{V R_T} (S_2 - S_1) - \frac{\Psi(\Delta \rho_2)}{\tilde{\omega} V R_T} S_2
\end{split}
\end{align}
with the gradients of the atmospheric temperature and salinity forcing $\Delta T_1^a \equiv T_{p1}^a - T_e^a$ and $\Delta S_1^a \equiv S_{p1}^a - S_e^a$. 
These are rewritten in terms of the heterogeneity of the two polar boxes 
$\Delta T_1^a \equiv \Delta T^a + \delta_T$ and $\Delta T_2^a \equiv \Delta T^a - \delta_T$, as well as $\Delta S_1^a \equiv \Delta S^a + \delta_S$ and $\Delta S_2^a \equiv \Delta S^a - \delta_S$. 
The temperatures and salinities are rescaled by $\alpha_T C$ and $\alpha_S C$, using the constant $C\equiv \sqrt{\frac{\gamma}{\tilde{\omega} V R_T}}$. Defining $\eta_1 \equiv \alpha_T C \Delta T^a$, $\eta_2 \equiv \alpha_S \frac{R_S}{R_T}C \Delta S^a$, 
$\eta_3 \equiv \frac{R_S}{R_T}$, $\omega \equiv 1 + \tilde{\omega}$, and $\kappa \equiv \frac{\tilde{\kappa}}{V R_T}$, yields
\begin{align}
\begin{split}
\dot{T}_1 &= \eta_1(1 + \delta_T) - T_1\left[ 1 + \kappa + \omega(T_1 - S_1)^2 \right] + T_2\left[ \kappa - (T_2-S_2)^2 \right]\\
\dot{S}_1 &= \eta_2(1 + \delta_S) - S_1\left[ \eta_3 + \kappa + \omega(T_1 - S_1)^2 \right] + S_2\left[ \kappa - (T_2-S_2)^2 \right]  \\
\dot{T}_2 &= \eta_1(1 - \delta_T) - T_2\left[ 1 + \kappa + \omega(T_2 - S_2)^2 \right] + T_1\left[ \kappa - (T_1-S_1)^2 \right]  \\
\dot{S}_2 &= \eta_2(1 - \delta_S) - S_2\left[ \eta_3 + \kappa + \omega(T_2 - S_2)^2 \right] + S_1\left[ \kappa - (T_1-S_1)^2 \right] .
\end{split}
\end{align}
The free parameters are $\eta_1$, $\eta_2$, $\eta_3$, $\kappa$, and $\omega$, as well as $\delta_T$ and $\delta_S$.
Two observables $q_{1,2} = T_{1,2} - S_{1,2}$ are defined, which are proportional to the meridional density gradient of the two boxes, and can be interpreted as representing the strength of the northward surface flow (although note that the actual flow strength $\Psi_{1,2}$ is proporional to the square of $q_{1,2}$).
The total northward flow into both boxes combined is defined as $q \equiv q_1 + q_2$.

When choosing non-zero heterogeneity $\delta_T$ or $\delta_S$ of the atmospheric forcing, besides the stable AMOC 'ON' ($q_1>0$ and $q_2>0$) and AMOC 'OFF' ($q_1<0$ and $q_2<0$) states, a stable state exists where the flow towards one of the polar boxes is reversed, i.e., with $q_1<0$ or $q_2<0$ but $q = q_1 + q_2 > 0$. This is then referred to as a partially collapsed ('PC') AMOC. It is a robust behavior of the model for a large range of parameter values, and the qualitative influence of the different parameters is discussed below.
A similar model was presented in \cite{NEF23}. Different here is the usage of a quadratic non-linearity, as well as the inclusion of coupling between the polar boxes and of dynamic temperature variables, which gives rise to an important dynamical regime not present in \cite{NEF23} (later referred as scenario II) and allows to assess EWS in a variety of observables.

For the remainder of this section, the dependence of the equilibrium solutions on the parameters are discussed. The main bifurcation parameter to induce the tipping point(s) will be $\eta_2$, which is proportional to the meridional gradient of the atmospheric salinity/freshwater forcing (averaged over the two polar boxes).
All other parameters will be fixed, and chosen as discussed. $\eta_1$ is proportional to the average atmospheric meridional temperature gradient, and besides quantitative differences, the behavior of the equilibria is the same when using $\eta_1$ as control parameter. It is beyond the scope of this paper to present all possible changes in equilibria in the space spanned by all seven parameters. What follows instead is a qualitative discussion of how each parameter affects the bifurcation diagram with respect to $\eta_2$. This yields two scenarios with different values of $\delta_T$, which cover all relevant qualitative scenario where a bifurcation towards a PC and OFF state occur.

For $\eta_1 \gtrsim 1$ (this was determined at small $|\delta_T|$ and $|\delta_S|$) the existence of PC is robust. The exact value of $\eta_1$ only shifts and stretches the bifurcation diagram with respect to $\eta_2$. Here, $\eta_1 = 3$ is chosen. $\eta_3$ represents the ratio of the salinity and temperature relaxation time scales. To get any multistability with coexisting ON and OFF states, $\eta_3$ needs to be sufficiently small such that the salt-advection feedback is dominant. This encodes the common assumption that the ocean temperature equilibrates faster to atmospheric anomalies than the ocean salinity. Here, $\eta_3 = 0.1$ is chosen.

$\omega$ should be clearly larger than 2 since the subtropical Atlantic is more voluminous than the polar basins. The exact choice of $\omega$ does not change the dynamics qualitatively. Larger $\omega$ shifts the bifurcation diagram w.r.t $\eta_2$ to higher values, since then larger atmospheric salinity gradients are needed to overcome the temperature-driven circulation. The value $\omega = 5$ is chosen. $\kappa$ is the strength of coupling of the two polar boxes. For larger values ($\kappa \gtrsim 0.5$) the possibility for a partially collapsed AMOC starts to disappear, as the densities in the two boxes become too tightly coupled. In this work, weak coupling with $\kappa = 0.2$ is assumed.

\begin{figure}
\includegraphics[width=0.99\textwidth]{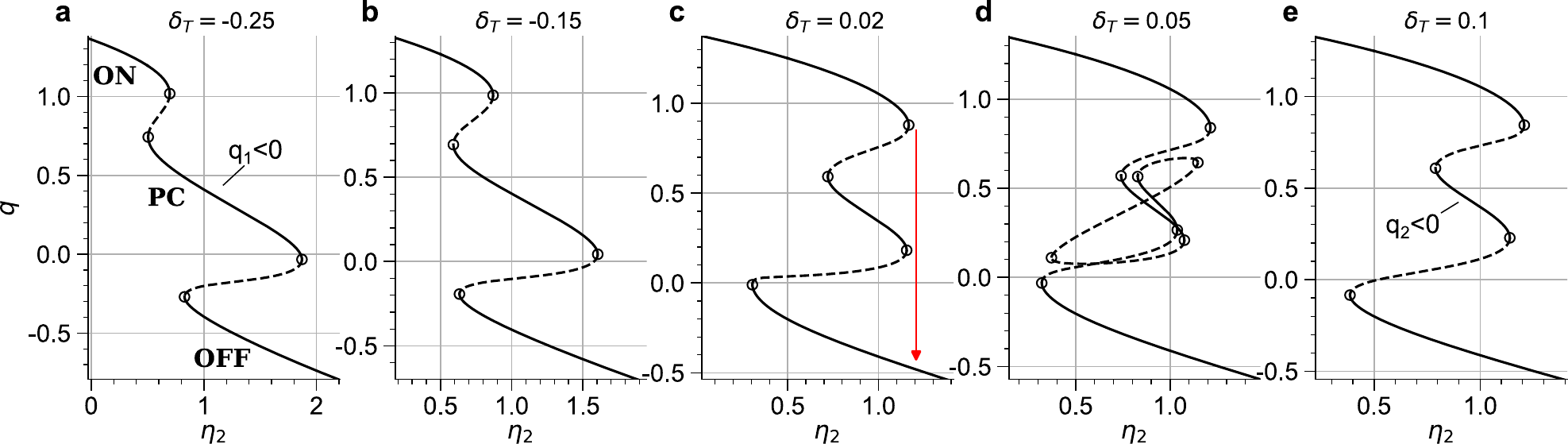}
\caption{\label{fig:bif_deltaT} 
Bifurcation diagrams of the box model with respect to the freshwater forcing parameter $\eta_2$, for five different values of the heterogeneity in surface temperature forcing $\delta_T$. Solid (dashed) lines are stable (unstable or saddle) fixed points. Open circles mark saddle-node bifurcations, or, as found in {\bf d}, fold bifurcations of two saddle points. 
}
\end{figure}

The parameters remaining to be fixed are the heterogeneities in freshwater and temperature forcing gradients $\delta_S$ and $\delta_T$. Without loss of generality it can be assumed that $\delta_S>0$, meaning polar box 1 is forced ``fresher'' than box 2. $\delta_S=0.1$ is chosen here.
The only remaining parameter that affects the qualitative behavior is $\delta_T$ (Fig.~\ref{fig:bif_deltaT}).
Depending on the sign of $\delta_T$, box 1 is either forced warmer or colder than box 2. If $\delta_T$ is above a certain magnitude and negative, warming and freshening compound each other and the bifurcation ON $\rightarrow$ PC happens already before the OFF branch exists (Fig.~\ref{fig:bif_deltaT}a). For intermediate, negative $\delta_T$, the three stable states can coexist (Fig.~\ref{fig:bif_deltaT}b) until a bifurcation at $\eta_2 \approx 0.7$ leaves only the OFF state and a PC state where the flow towards box 1 reversed (i.e., $q_1<0$, although this is not visible from the bifurcation diagram here). When increasing $\delta_T$ further towards 0, the extent of the ON branch increases. As soon as $\delta_T>0$, the implied stronger cooling of box 1 partially counteracts the effect on density of the freshening by $\delta_S>0$. For small $|\delta_T|$ box 1 is still less stable, but now the AMOC ON branch extends further than the partially collapsed branch, i.e., the ON branch collapses directly onto the OFF branch upon increasing $\eta_2$ (Fig.~\ref{fig:bif_deltaT}c).
Upon slight increase of $\delta_T$, the heterogeneities in salinity and temperature almost perfectly cancel each other. This yields a regime with four stable coexisting fixed points, featuring two partially collapsed states of either $q_1<0$ or $q_2<0$, along with a complicated sequence of bifurcations and three additional unstable branches (Fig.~\ref{fig:bif_deltaT}d). Importantly, the ON branch still extends to higher $\eta_2$ compared to the partially collapsed branches. When $\delta_T \gtrsim 0.075$ it overrides $\delta_S>0$ and there is again only one partially collapsed state, but now with $q_2<0$ while $q_1>0$. When $|\delta_T|$ is still small enough the ON branch still extends furthest (Fig.~\ref{fig:bif_deltaT}e). Thereafter, increasing $\delta_T$ decreases the extent of the ON branch to lower $\eta_2$, and the bifurcation diagram becomes equivalent to Fig.~\ref{fig:bif_deltaT}b, but with the roles of box 1 and 2 exchanged. 

\begin{figure}
\includegraphics[width=0.5\textwidth]{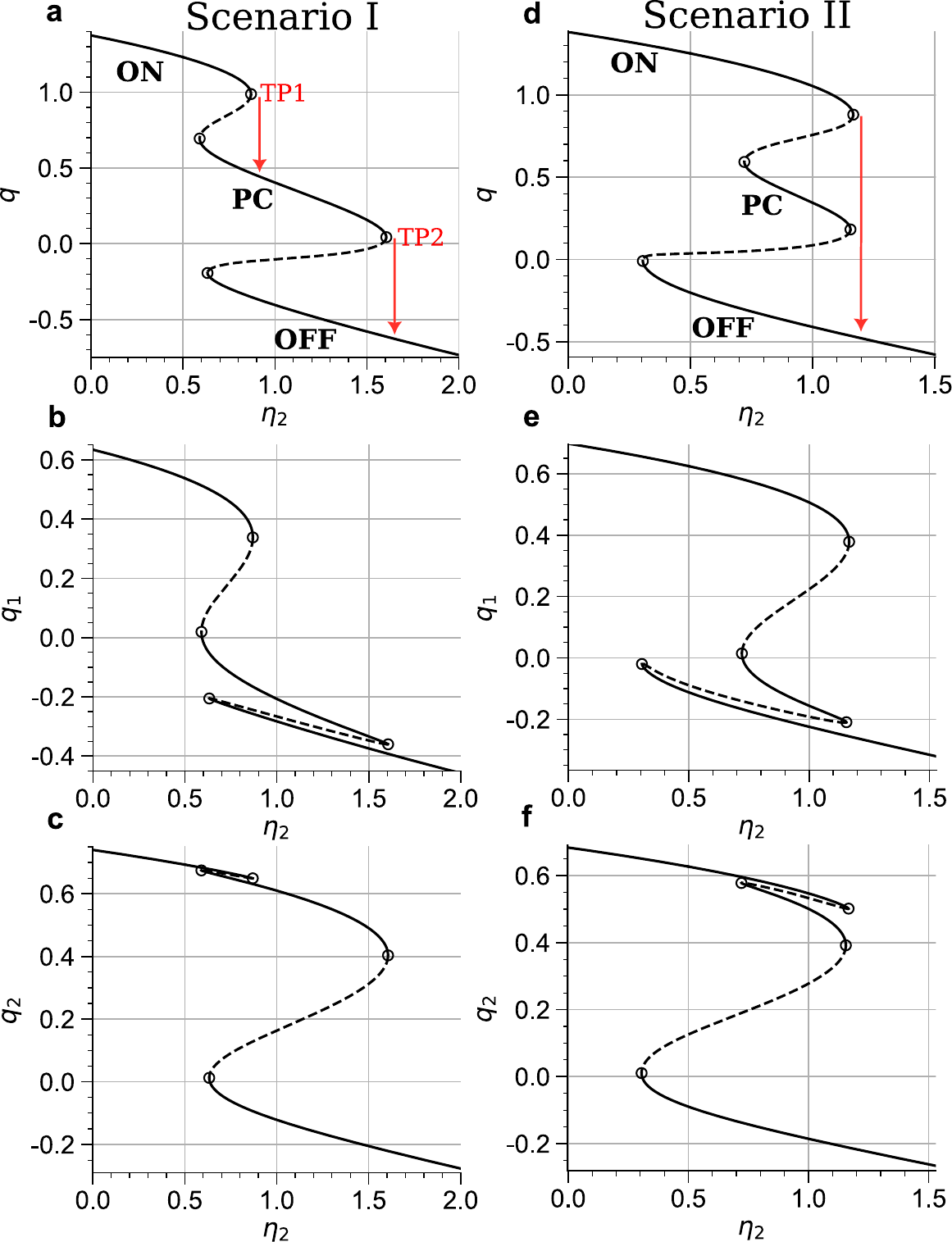}
\caption{\label{fig:bif_eta2} 
Bifurcation diagrams with respect to the freshwater forcing parameter $\eta_2$ for the three observables $q$, $q_1$, and $q_2$. For scenario I ({\bf a-c}) $\delta_T = -0.15$ is chosen, i.e., box 1 is forced fresher and warmer than box 2. In scenario II ({\bf d-f}) $\delta_T = 0.02$ is chosen, such that box 1 is forced fresher but slightly cooler. 
}
\end{figure}

From the previous paragraph it follows that there remain two qualitative scenarios, as far as sequences of bifurcations starting from the ON state are concerned.
{\it Scenario I:} One of the polar boxes is forced fresher and warmer. When increasing $\eta_2$ there is an intermediate TP to a partially collapsed state before the full collapse to OFF.
This is the case for Fig.~\ref{fig:bif_deltaT}a,b.
{\it Scenario II:} The atmospheric temperature and salinity forcings partly counteract such that a partially collapsed state still exists but the AMOC collapse is directly from the ON to the OFF state. 
This is the case for Fig.~\ref{fig:bif_deltaT}c,d, and
also Fig.~\ref{fig:bif_deltaT}e with exchanged roles of box 1 and 2.
The remainder of the paper focuses on these two scenarios, by choosing $\delta_T = -0.15$ (Fig.~\ref{fig:bif_deltaT}b) and $\delta_T = 0.02$ (Fig.~\ref{fig:bif_deltaT}c).
Bifurcation diagrams for the two scenarios projected onto the observables $q$, $q_1$, and $q_2$ are shown in Fig.~\ref{fig:bif_eta2}. In scenario I (Fig.~\ref{fig:bif_eta2}a-c), when increasing $\eta_2$ on the ON branch there is a first bifurcation (called TP1 hereafter) where the convection and circulation in box 1 collapses, and thus $q_1$ becomes negative while $q_2$ barely changes.
Further increasing $\eta_2$ on the PC branch, the circulation to box 2 collapses at the second bifurcation (TP2) where $q_2$ and $q$ become negative. For scenario II (Fig.~\ref{fig:bif_eta2}d-f), increasing $\eta_2$ on the ON branch can only lead to a collapse directly from ON to OFF. In the next section it is shown that statistical EWS may display a variety of trends when measured from different observables and for the different bifurcations (TP1 and TP2) and scenarios.

\section{Statistical early-warning signals}
\label{sec:ews}

\begin{figure}
\includegraphics[width=0.42\textwidth]{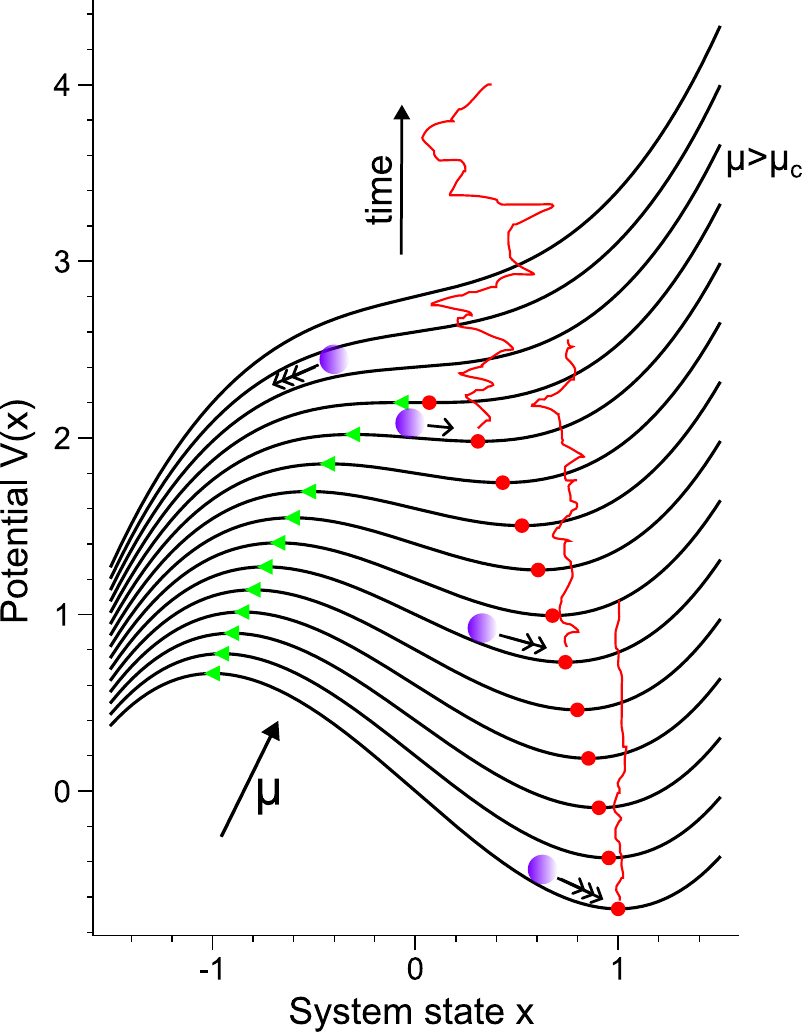}
\caption{\label{fig:csd}
Changing potential landscape $V_{\mu}(x)$ for a one-dimensional system described by the saddle-node normal form $\dot{x} = -\partial_x V_{\mu}(x) = x^2 + \mu$. As the control parameter $\mu$ is increased, the stable node (red) and the saddle point (green triangles) approach each other, and at the critical value $\mu_c = 0$ they collide and no more fixed point exists.
If the system is perturbed away from the stable node, it takes increasingly long to relax back (purple balls). When driven by small, random perturbations, the fluctuations around the stable node become increasingly large and correlated (see red timeseries).
}
\end{figure}

The first paragraph in this section gives a quick pedagogical recap of EWS for a general system with state variable $x$, whereafter EWS in the conceptual AMOC model is discussed. Before a saddle-node bifurcation, the relaxation dynamics towards the stable fixed point become very slow along one particular degree of freedom.
This is CSD and can be understood as a flattening of the potential $V_{\mu}(x)$ (more generally quasipotential \cite{Graham1991}) governing the dynamics $\dot{x} = -\partial_x V_{\mu}(x)$, as shown for one dimension in Fig.~\ref{fig:csd}.
Far from the TP, the potential well is steep, and thus the relaxation towards the stable fixed point after a perturbation is very fast. The well gets more shallow and flat as the bifurcation is approached, and the relaxation becomes slower. Shortly before the TP, the potential is almost perfectly flat, and the relaxation becomes arbitrarily slow with a restoring rate $\lambda$ going to zero.
From data, CSD may be detected if there are random perturbations from a system's environment, such as small-scale atmospheric anomalies that perturb the ocean circulation. Their effect will grow towards the TP (red trajectories in Fig.~\ref{fig:csd}), since in a flatter potential perturbations of a given amplitude push the system further away from the fixed point. One can thus measure increases in the variance and autocorrelation of the fluctuations around the slowly evolving mean state in time series. These are statistical EWS.

\begin{figure}
\includegraphics[width=0.7\textwidth]{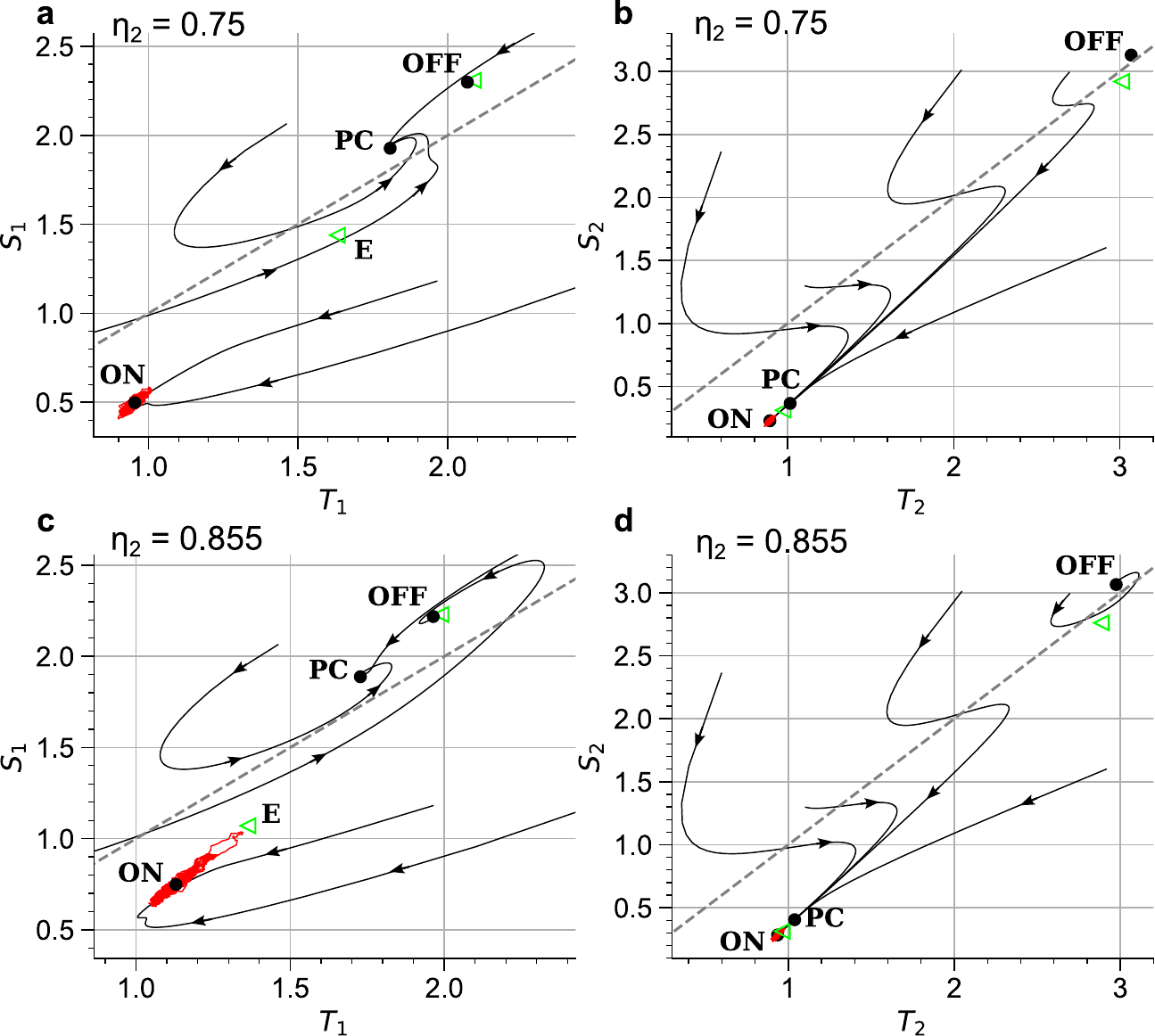}
\caption{\label{fig:phaseportrait} 
Projected phase portraits of the AMOC box model (for scenario I with $\delta_T=-0.15$) for two values of the freshwater forcing $\eta_2 = 0.75$ ({\bf a,b}) and $\eta_2 = 0.855$ ({\bf c,d}), the latter being close to the bifurcation with partial AMOC collapse. The stable fixed points are marked with dots, and the saddle points with open triangles.
In red are trajectories driven by isotropic noise with $\sigma=0.03$. A few simulations without noise forcing (black lines and arrows) indicate the deterministic flow.
}
\end{figure}

For saddle-node bifurcations in higher-dimensional systems the picture of Fig.~\ref{fig:csd} applies {\it locally} in phase space when close to the bifurcation due to the center manifold theorem. Here, one degree of freedom becomes by far the slowest, and after a short transient the system becomes effectively one-dimensional and confined to the degree of freedom that lives on the (extended) center manifold. 
For the AMOC box model this is demonstrated in Fig.~\ref{fig:phaseportrait}, where projections of phase space onto the variables ($T_1$, $S_1$) and ($T_2$, $S_2$) are given at two parameter values, one relatively far from (Fig.~\ref{fig:phaseportrait}a,b) and one close to the TP (Fig.~\ref{fig:phaseportrait}c,d). 
In red are single trajectories initialized in the ON state, driven by additive isotropic noise, whereby the state is then governed by the stochastic differential equation 
\begin{equation}
 dT_{1,t} = \left(\eta_1(1 + \delta_T) - T_{1,t}\left[ 1 + \kappa + \omega(T_{1,t} - S_{1,t})^2 \right] + T_{2,t}\left[ \kappa - (T_{2,t}-S_{2,t})^2 \right] \right)dt + \sigma dW_{T_1,t}
\end{equation}
with a standard Wiener process $W_{T_1,t}$, and equivalently for the other variables with individual noise processes. 
Here and elsewhere, simulations are performed with an Euler-Maruyama scheme with time step 
$dt = 0.005$ and noise strength $\sigma=0.03$. When close to the TP (panels c,d) the noise-driven dynamics (red lines) becomes stretched along a degree of freedom that is close to a line with $T_1 - S_1 = \text{const}$ (gray dashed lines are $T_1 = S_1$ in panels a,c and $T_2 = S_2$ in c,d). This degree of freedom is typically directed towards the saddle point \cite{LOH25} (marked 'E' in Fig.~\ref{fig:phaseportrait}), also known as the edge state, which collides with the ON state at the bifurcation.
The fluctuations in the ($T_2$, $S_2$)-projection do not grow significantly, since irrespective of $\eta_2$ the ON, PC and edge states are very closeby in ($T_2$, $S_2$) and thus barely move as the TP is approached. 
For TP2 from the PC to the OFF state the stretching of fluctuations is in a direction close to $T_2 - S_2 = \text{const}$ (which points towards the other saddle point) with only little change in the ($T_1$, $S_1$)-plane. In Scenario II, where tipping is directly from ON to OFF, the phase portrait is very similar to Fig.~\ref{fig:phaseportrait} and the increasing fluctuations are again stretched predominantely in the ($T_1$, $S_1$)-plane
along a similar direction towards the edge state. 

\begin{figure}
\includegraphics[width=0.75\textwidth]{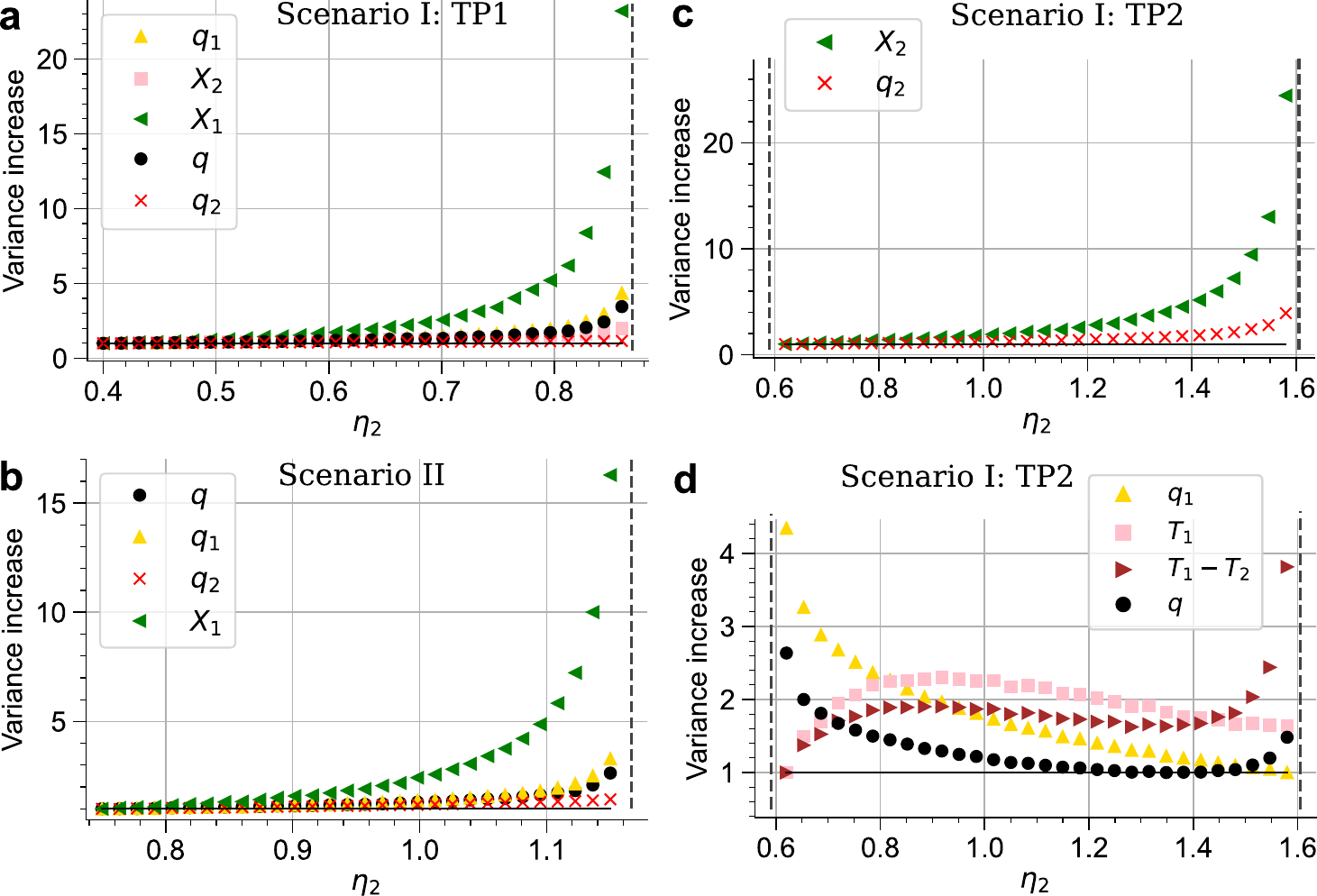}
\caption{\label{fig:ews} 
Changes in variance as $\eta_2$ is increased towards the tipping points in the conceptual AMOC model, measured for different observables, including $X_{1,2} = T_{1,2} + S_{1,2}$. The bifurcations are marked by vertical dashed lines. {\bf a} Change of $\eta_2$ towards the partial AMOC collapse (TP1) for scenario I. {\bf b} Change of $\eta_2$ towards the full AMOC collapse in scenario II. {\bf c,d} Changes in variance on the PC branch (scenario I), which is bounded by a bifurcation on either side.
The variance has been obtained by ensembles of 100 noise-driven simulations ($\sigma=0.03$)
of duration $t=100$ each, initialized in the ON ({\bf a,b}) and PC ({\bf c,d}) fixed points at the respective fixed value of $\eta_2$. When close to the bifurcations, there are rare noise-induced transitions, and the corresponding realizations are discarded. The variance shown for each observable has been normalized by its minimum value on the respective branch. 
}
\end{figure}

EWS are masked in certain observables because of the clear preferred direction of large fluctuations towards the edge state, which is almost parallel to lines with $T_1 - S_1 = \text{const}$. Hence, increases in noise-driven fluctuations are not seen much in the observable $q_1 = T_1 - S_1$ (Fig.~\ref{fig:ews}a). Only when very close to the tipping point the variance rises, since the level sets of $q_1$ are not perfectly parallel to the direction of the edge state. The observable $q$ shows a similarly weak signal, and for observables restricted to the ($T_2$, $S_2$)-plane almost no increase in variance is seen. 
In contrast, the observable $X_1 \equiv T_1 + S_1$ is aligned with the preferred direction and shows very strong increases in variability. 
This is consistent with findings from the global ocean model Veros \cite{LOH24c,LOH25}, where the edge state and EWS were also aligned with Atlantic ``spiciness'', i.e., $T+S$, and not density gradients or AMOC strength, i.e. the orthogonal $T-S$. 

It may seem unsurprising that for anticipating the collapse of box 1 the variables related to box 1 are better than those related to box 2. One may even argue that the difference in EWS allows one to determine which aspect of the circulation will undergo tipping. 
This is, however, not always possible, as becomes evident in scenario II, where tipping is directly from ON to OFF (Fig.~\ref{fig:ews}b). Here, besides the clear dependence on the choice of observable (e.g. $X_1$ versus $q_1$ or $q$), using variables only from box 2 would still result in insignificant variance (or autocorrelation) increase and thus a missed warning of a collapse of the circulation to box 2. 
The interpretation of EWS is blurred further, since more complicated, non-monotonic behavior of EWS can occur, as exemplified by TP2 from PC to OFF in scenario I (Fig.~\ref{fig:ews}c,d). Here, tipping occurs from a branch that is bounded on either side by a bifurcation. Thus, CSD and EWS should be seen both for increasing and for decreasing $\eta_2$. Indeed, observables such as $q$ show this behavior (Fig.~\ref{fig:ews}d), reaching minimum variance only shortly before the collapse from PC to OFF. Thus, when starting to measure variability around a baseline state from an arbitrary value of the control parameter, one cannot necessarily expect EWS to increase monotonically towards the TP, as was also discussed in \cite{LOH24} (see Fig.~8 therein). 

For the conceptual model considered here this only happens on the PC branch and not on the ON branch serving as analogue of the present-day AMOC. But in general we cannot exclude that in the past centuries the Atlantic ocean circulation has experienced bifurcations, and hence that the present-day state is also bounded by a bifurcation going back in time (at reduced anthropogenic forcing). Multiple bifurcations (intermediate tipping points \cite{LOH24}) might have and may still happen towards collapse. These may not concern qualitative changes in convection, but changes in the stable spatio-temporal pattern of the Atlantic ocean circulation \cite{LOH24}. 

Further types of EWS trends on the PC branch are seen in other observables (Fig.~\ref{fig:ews}d). The variability may decrease towards the bifurcation on either side (e.g. $T_1$), or there can be an increase in variability towards one bifurcation and a decrease towards the other, with a monotonic (e.g. $q_1$) or non-monotonic (e.g. $T_1-T_2$) trend in between, meaning there may be temporary increases in variability without immediate approach of a TP.

\section{Prediction of tipping times by extrapolation} 
\label{sec:prediction}

The previous section showed that in principle - depending on the observable - any possible trend in EWS may be seen when approaching a TP. Next it is shown that this leads to large uncertainties and biases when extrapolating a trend in variance or autocorrelation to predict the exact time when tipping will occur.

Before returning to the AMOC box model, in this paragraph it is explained how for a general system this extrapolation can be done by exploiting that towards a saddle-node bifurcation the restoring rate $\lambda$ in the critical degree of freedom goes to zero. Importantly, three conditions need to be met. First, $\lambda$ needs to be reconstructed from statistical EWS in some observable. Second, the observable needs to project well on the critical degree of freedom.
Third, it needs to be known how $\lambda$ scales as a function of the control parameter $\mu$, such that one can extrapolate to where $\lambda$ reaches 0 as $\mu$ crosses the critical value. 
Previous studies, which predicted tipping of the Greenland ice sheet \cite{BOE21} and the AMOC \cite{DIT23} from data, implicitly assumed these three conditions to hold by positing that whatever observable is measured should obey the saddle-node normal form, i.e., obey the one-dimensional dynamics of the critical degree of freedom on the extended center manifold, and that it did so for the entire period of historical observations.
The saddle-node normal form (with additive noise) is given by
\begin{equation}
dx_t = (x_t^2 - \mu)dt + \sigma dW_t, 
\end{equation}
where $\mu = 0$ demarcates the bifurcation. The system can be linearized around the stable fixed point $x* = -\sqrt{\mu}$, and approximated by the Ornstein-Uhlenbeck process $d\tilde{x}_t = -\lambda \tilde{x}_t dt + \sigma dW_t$. 
The linear restoring rate is $\lambda = -\partial_x (x^2 - \mu)|_{x=x*} = 2 \sqrt{\mu}$. 
This linearized, reversible stochastic process can be discretized, which gives the desired link of the linear restoring rate and statistical EWS as measured from discrete time series samples. Specifically, data sampled at small time intervals $\Delta_t$ can be approximated by an AR(1) process
\begin{equation}
X_{k+1} = e^{-\lambda \Delta t} X_k + \epsilon_k,
\end{equation}
where $\epsilon_k$ are Gaussian random variables with variance $\sigma^2 (2\lambda)^{-1} (1-e^{-2\lambda \Delta t})$. For this process the lag-1 autocorrelation is given by $\rho_1 = e^{- \lambda \Delta t}$. Since for the normal form $\lambda = 2 \sqrt{\mu}$, we can reconstruct $\mu$ from data by
\begin{equation}
\label{eq:ctrl_param}
\mu = \lambda^2/2 = \left( \frac{\ln \rho_1}{2 \Delta t} \right)^2. 
\end{equation}
Thus, for a single time series one can estimate $\rho_1$ in a sliding window as a function of time, and, assuming a linear trend in $\mu$, estimate with a linear fit to the function on the righthand-side at what time $\mu$ will cross zero, which is exactly when the autocorrelation tends to 1 at the saddle-node bifurcation. 

\begin{figure}
\includegraphics[width=0.45\textwidth]{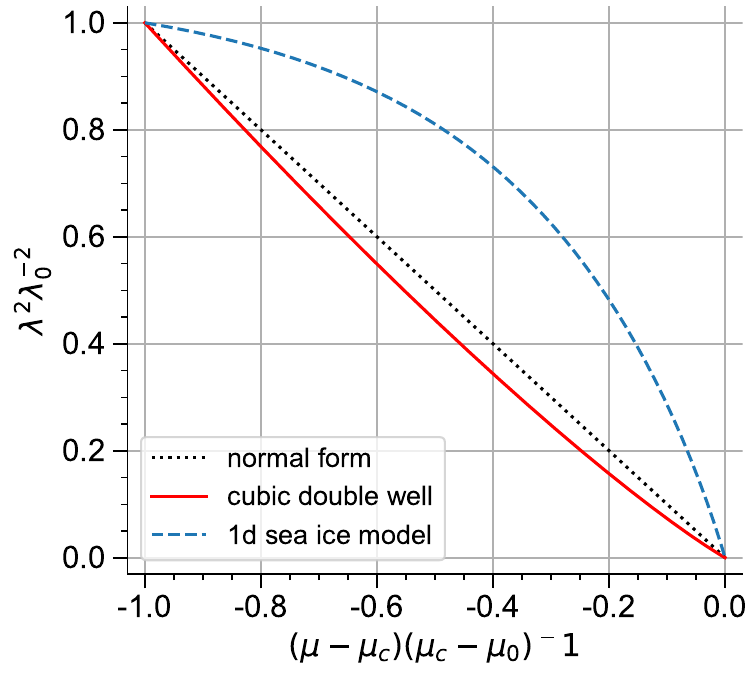}
\caption{\label{fig:relaxation_rates} 
Scaling of the square of the linear restoring rate $\lambda$, normalized by its value $\lambda_0$ at an arbitrary starting value $\mu_0$ of the control parameter, as function of the control parameter $\mu$ for three one-dimensional systems. The horizonal axis is the control parameter scaled by the distance from $\mu_0$. For the cubic double-well model $\mu_0=-0.4$ is chosen, and the bifurcation is at $\mu_c \approx 0.383$. The sea ice model is given by $\dot{I} = \Delta \tanh (I/h) - BI + L - F - 1 + \mu$, with parameter values $\Delta = 0.43$, $h=0.5$, $B=0.45$, $L=1.25$, and $F=1/25$. The fixed point is computed with the Newton method. The starting value is $\mu_0=-0.4$ and the bifurcation is at $\mu_c \approx -0.1084$.
}
\end{figure}

However, even for univariate systems, the normal form and associated scaling holds only when close to the bifurcation, and not if the data extends further away from the bifurcation.
In Fig.~\ref{fig:relaxation_rates} the normal form scaling $\mu \propto \lambda^2$ (dotted line) is compared to the scaling of two non-normal form examples. For the simple double-well potential $\dot{x} = x - x^3 + \mu$ (solid red line) the function $\lambda^2 (\mu)$ is concave, as pointed out in \cite{LOH25b}, and thus a linear extrapolation will always predict the bifurcation too early. 
Similarly, as pointed out in \cite{BEN24}, data from the Stommel-Cessi model \cite{CES94} gives a too early prediction. In other cases the function can be convex, for instance when the non-linearity is a $\tanh$-function as in the one-dimensional sea ice model from \cite{EIS09}, which is shown in Fig.~\ref{fig:relaxation_rates} in a simplified version without seasonal cycle (as in \cite{LOH21b}). In this case, the prediction by linear extrapolation (i.e. assuming the normal form) will always be too late.

\begin{figure}
\includegraphics[width=0.95\textwidth]{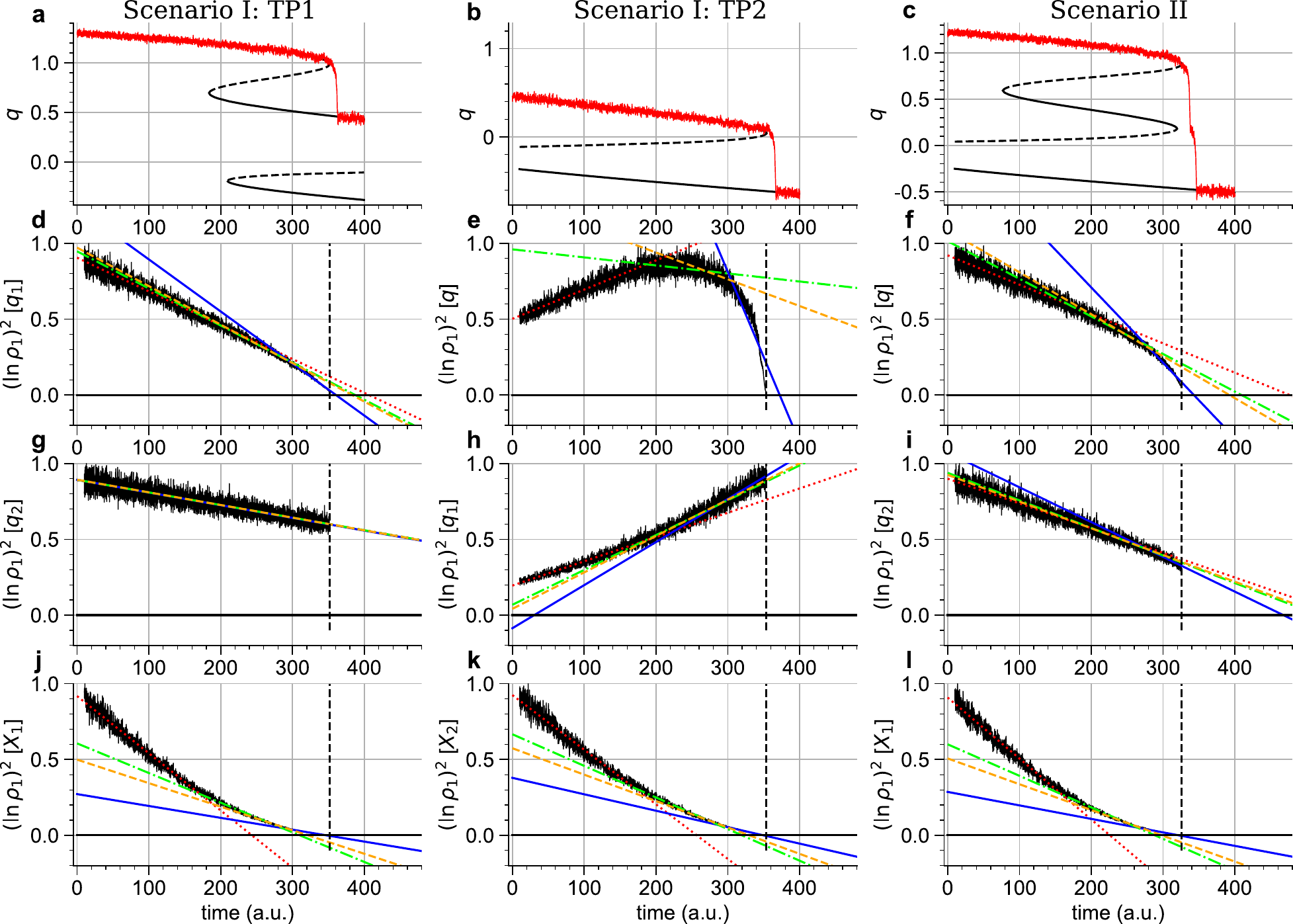}
\caption{\label{fig:tipping_time} 
Extrapolation of the lag-1 autocorrelation $\rho_1$ to estimate the time of tipping, at which the linear restoring rate $\lambda \propto (\ln \rho_1)$ of the critical degree of freedom reaches 0. Simulations are performed with a linear increase in $\eta_2$ over time. The left and middle column show model scenario I with tipping from ON to PC ($\eta_2$ ramped from 0.3 to 0.95), and PC to OFF ($\eta_2$ from 0.9 to 1.7), respectively. The right column is model scenario II, i.e., a direct tipping from ON to OFF ($\eta_2$ from 0.6 to 1.3). Panels {\bf a-c} show an example realization overlaid on the bifurcation diagram. The remaining panels show the evolution of the quantity $(\ln \rho_1)^2$, where $\rho_1$ was computed at each time point from an ensemble of $N=7000$ realizations ($\sigma = 0.03$) for each of the three tipping points. 
$\rho_1$ is computed until a cutoff time where 99\% of the ensemble members have not tipped yet. The 1\% of realizations that tip are discarded. The vertical dashed line marks the time where the bifurcation is crossed. The normal form assumption predicts a linear decrease of $(\ln \rho_1)^2$, and the remaining lines are linear fits to different portions of the time series (dotted line is a fit to the first half, and the solid line to the final sixth). 
}
\end{figure}

This effect gets amplified in multivariate system like the AMOC box model, as shown in Fig.~\ref{fig:tipping_time}. Here, for many variables and observables the normal form assumption only becomes tenable when arbitrarily close to the bifurcation. For tipping from ON to PC (first column), $\lambda^2$ estimated from $q_1$ is convex (Fig.~\ref{fig:tipping_time}d), yielding late predictions although the slope deviates only slightly from the normal form expectation, which is likely coincidental for this observable in this model. Other observables, such as $q_2$ (Fig.~\ref{fig:tipping_time}g), show a consistently linear, but too shallow slope in $\lambda^2$. Using the observable $X_1 = T_1 + S_1$ yields a concave function for $\lambda^2$ (Fig.~\ref{fig:tipping_time}j), initially giving a quite early estimate. The estimate rapidly improves as data closer to the bifurcation becomes available.
Importantly, because $X_1$ projects very well on the critical degree of freedom and gives strong EWS, the signal-to-noise ratio in $\lambda^2$ is higher compared to other observables, which in practice gives a more certain (although still biased) tipping estimate. 
Note that in general when $\lambda$ is estimated from data, it may initially be set by another slow degree of freedom that is not the critical one \cite{LOH25b,ASH25}, in which case the correct scaling only emerges shortly before the bifurcation. 

For tipping from PC to OFF (second column) more diverse trends can be seen. Since the PC branch is bounded by a bifurcation on either side, some observables, such as $q$ (Fig.~\ref{fig:tipping_time}e), show an increasing trend in $\lambda^2$ which is only reversed shortly before the TP. The trend of $\lambda^2$ in observables showing a decrease in variability all the way to the bifurcation, such as $q_1$ (Fig.~\ref{fig:tipping_time}h), consequently has the wrong sign. The observable $X_2 = T_2 + S_2$ (related to the edge state direction) becomes eventually accurate, although it is again very concave (Fig.~\ref{fig:tipping_time}k). For scenario II (third column), $\lambda^2$ estimated from $q$ is concave and up until the bifurcation gives a substantial overestimate of the tipping time. $q_2$ fails to give a warning, even though the entire circulation collapses. The observable $X_1$ is again best in terms of accuracy and precision (low variance in $\lambda^2$), even though $\lambda^2$ is strongly concave.

\section{Discussion}
\label{sec:discussion}

This paper presents a conceptual model as a minimal example for how spatially distributed deep convection leads to different possible pathways of an AMOC collapse that depend on the heterogeneity of the atmospheric forcing, and how from EWS alone it is difficult to assess when a tipping will occur and what state it will lead to (partial or full AMOC collapse). The following was demonstrated: 

\begin{enumerate}
 \item Existence and strength of EWS is observable-dependent. Observables that seem obvious (as shown here for the AMOC strength represented by $q$) can perform poorly. Physical knowledge \cite{VWE24,SMO25} or the design of observables from dynamical systems principles (edge states \cite{LOH25}) or operator-theoretic arguments \cite{LUC24,LOH25b} is required.
 
 \item The AMOC as a spatially extended, heterogeneous system may display higher multistability. The simplest case is tri-stability due to two convection sites, as shown here and similarly in \cite{NEF23}. 
 
 \item Multistability can lead to intermediate TPs \cite{LOH24} with partial AMOC collapse. However, as in scenario II, an AMOC collapse can also bypass partially collapsed states. EWS signify loss of {\it local} stability (in phase space) and, unless special observables are known beforehand, lack global information to indicate whether the collapse is full or partial. 
 
 \item A given branch of stable states can be bounded on either side by a TP, resulting in non-monotonic EWS. This cannot be excluded for the present-day AMOC. 
 
 \item The normal form assumption, used for extrapolating EWS towards the expected time of tipping, may only hold when measuring (unknown) special observables or being very close to the TP. 
 Even if one knows roughly which variables are relevant (e.g. box 1), different observables give different estimates of the time of tipping. This structural statistical uncertainty on the estimate may be unbounded unless there are arguments to exclude certain types of observables. 
\end{enumerate}


The model is not intended to inform about the likelihood of future AMOC scenarios, or to argue for or against certain observables to be used as EWS. It serves to illustrate how going up one step in complexity from Stommel's model already poses issues for the interpretation of EWS, should they be observed in data. While there is clear evidence for heterogeneity in atmospheric forcing \cite{BAM12,CHA19}, it is not argued here whether scenario I or II is more likely in reality, since this depends on the interpretation of the model boundary conditions and the definition of box 1 and box 2. 

AMOC complexity and associated challenges for the interpretation of EWS are plausibly further enhanced for several reasons. First, the array of convection sites that can collapse non-simultaneously could be further fragmented, noting, e.g., the different strenghts and mechanisms of deep-water formation in the Labrador and Irminger seas \cite{PIC03,PET20}. 
Second, the presented model only encodes the salt-advection feedback, whereas other positive feedbacks such as the convective \cite{WEL82} and subpolar gyre feedbacks \cite{BOR13} may be present, which could yield further pathways to a partial or full AMOC collapse that are difficult to distinguish from EWS and naive observables. 

Third, instead of collapsing, deep water formation could migrate \cite{ART23}. It has been argued that downwelling in the Atlantic cannot fully collapse unless it is replaced by downwelling elsewhere to balance the (unchanged) Southern ocean wind-driven upwelling \cite{BAK25}. This could be by development of a Pacific meridional overturning, which in current climate models only happens to some degree under extreme global warming \cite{BAK25}. 
Alternatively, deep water formation may move northwards, and there is observational support for this due to ``Atlantification'' of the Arctic \cite{ART25}. This could represent tipping to a new state, comprising abrupt sea ice change and the onset of convection at new sites, or a more gradual process acting as stabilizing feedback to prevent an AMOC collapse. While CMIP6 models nevertheless show transitions to states with very weak AMOC and only rarely develop Arctic deep convection \cite{HEU24}, they may not represent the Atlantification well enough yet \cite{MUI23}. 


Further factors that are not fully accounted for in most model projections may influence the tipping behaviour of the AMOC, and thus complicate the interpretation of EWS. Increased Antarctic meltwater could weaken the effect of Greenland melting on the AMOC \cite{SIN25}. An AMOC slowdown
may also reduce freshwater forcing from Greenland due to the decrease in heat transport \cite{POP25}. Further, Indian ocean warming may provide a stabilizing feedback for the AMOC \cite{HU19}. 
Taking together the various alternative feedbacks and possible alternative configurations of the AMOC mentioned above, it is not clear whether the AMOC collapses in CMIP6 models under freshwater and global warming forcing can be attributed to a singular TP due to the salt-advection feedback alone. As a result, it is unclear what kind of transition the detection of EWS in any real-world observable would indicate. 

A recent study does provide an argument as to why the mean AMOC strength in a CMIP6-class model should approximately scale as the square-root of the freshwater forcing acting as control parameter \cite{VWE25c}, which would be the hallmark (though not necessarily proof) of a saddle-node normal form. However, it is unclear whether it is consistent that the same observable in this model does not display CSD. It would be interesting to investigate whether other models support the scaling, and whether other arguments and approximations could support a different scaling, also when considering temperature as control parameter. 
It would also be helpful to clarify to what degree state-of-the-art models feature deterministic chaotic oscillations in the ocean dynamics (e.g. Atlantic multidecadal variability not driven by atmospheric variability), which would challenge the saddle-node bifurcation picture. Under chaotic oscillations the AMOC collapse TP would be a {\it boundary crisis} \cite{OTT93}, which may still be associated with CSD, but one needs to take care in finding robust observables for EWS in this case \cite{LOH25b}, especially since it is a priori unknown how oscillatory modes will change as an AMOC collapse is approached \cite{LOH24}. 


In conclusion, EWS are an appealing option to complement the assessment of the risk and critical forcing for an AMOC collapse, given current observational and modeling uncertainties. Their interpretability is, however, limited since it is unclear which observables are robust and representative, and because they do not indicate what system state is reached at the bifurcation. One may attempt to rationalize which aspect of the circulation will collapse based on the difference in EWS of various observables, but there may nevertheless be situations (as in the scenario II here) where this can fail. The non-normal form behavior of observables and possible non-monotonicity of EWS challenges the possibility for quantitative prediction of an AMOC collapse via extrapolation of EWS.

\bibliography{refs} 



\end{document}